\documentclass[12pt]{article}
\usepackage{amsmath}
\usepackage{mathtools} 
\usepackage[T1]{fontenc}
\usepackage{amssymb}
\usepackage{amsfonts}
\usepackage{amsthm}
\usepackage{mathrsfs}
\usepackage[all]{xy}
\usepackage{tensor}
\usepackage{comment}
\usepackage{stmaryrd}
\usepackage{bm}
\usepackage[normalem]{ulem}
\usepackage[usenames,dvipsnames]{xcolor}
\usepackage[colorlinks=true, citecolor=Blue, linkcolor=Blue]{hyperref}
\usepackage{ragged2e}
\usepackage{tikz}
    \usetikzlibrary{decorations.markings}

\usepackage{diagbox}  
\usepackage{makecell} 
    
\usepackage[labelfont={small, bf, sf}, textfont={small,sf}]{caption}
\numberwithin{equation}{section} 
\usepackage{authblk}
\usepackage[in]{fullpage}
\usepackage{cite}

\DeclareRobustCommand{\rchi}{{\mathpalette\irchi\relax}}
\newcommand{\irchi}[2]{\raisebox{\depth}{$#1\chi$}} 

\newcommand{\clp}{\text{CLP}}
\newcommand{\Amb}{A}
\newcommand{\ext}{\text{ex}}

\newcommand{\vep}{\varepsilon} 

\newcommand{\lie}{\pounds}

\newcommand{\brmod}[2]{\llbracket #1, #2 \rrbracket}

\newcommand{\h}[1]{{\hat{#1}}}
\newcommand{\wh}[1]{{\widehat{#1}}} 
\newcommand{\fs}{\mathscr{F}}

\newcommand{\beom}{\mathcal{E}} 
\newcommand{\stm}{\mathcal{M}}
\newcommand{\stmr}{\stm_r}
\newcommand{\sr}{\mathcal{U}}
\newcommand{\ns}{\mathcal{N}}

\newcommand{\flx}{\mathcal{F}}
\newcommand{\n}[1]{\mathscr{#1}}

\newcommand{\ra}{\rightarrow}

\newcommand{\p}[1]{  {\underline{ #1} }   }

\newcommand{\beq}{\begin{equation}}
\newcommand{\eeq}{\end{equation}}
\newcommand{\bes}{\begin{subequations}}
\newcommand{\ees}{\end{subequations}}
\newcommand{\bea}{\begin{eqnarray}}
\newcommand{\eea}{\end{eqnarray}}

\newcommand{\be}{\begin{equation}}
\newcommand{\ee}{\end{equation}}

\title{Ambiguity resolution for integrable gravitational charges}

\author[1]{Antony J. Speranza\thanks{asperanz@gmail.com}}

\affil[1]{\small \it Department of Physics, University of Illinois, Urbana-Champaign, Urbana IL 61801, USA}

\date{January 31, 2022}

\begin{document}

\maketitle

\begin{abstract}

Recently, Ciambelli, Leigh, and Pai (CLP)
[arXiv:2111.13181]
have shown that nonzero  charges 
integrating Hamilton's equation can be defined for all diffeomorphisms
acting near the boundary of a subregion in a gravitational theory.  
This is done by 
extending the phase space to include a set of 
embedding fields that parameterize the location of the 
boundary.  Because their construction differs from previous
works on extended phase spaces by 
a covariant phase space ambiguity, the question arises 
as to whether the resulting charges are unambiguously defined.
Here, we demonstrate that ambiguity-free charges can be obtained 
by appealing to the variational principle for the subregion,
following recent developments on dealing with boundaries in the 
covariant phase space.  Resolving the ambiguity produces 
corrections to the diffeomorphism charges, and also generates 
additional obstructions to integrability of Hamilton's equation.  
We emphasize the fact that the CLP extended phase space produces
nonzero diffeomorphism charges distinguishes it from previous constructions
in which diffeomorphisms are pure gauge, since the embedding fields 
can always be eliminated from the latter by a choice of unitary gauge. 
Finally, we show that Wald-Zoupas charges, with their characteristic 
obstruction to integrability, are associated with a modified 
transformation in the extended phase space, clarifying the 
reason behind integrability of Hamilton's equation for standard 
diffeomorphisms.

\end{abstract}

\flushbottom

\newpage

\section{Introduction}

Hamiltonian charges associated to diffeomorphisms constitute an important 
set of observables in gravitational theories.  While bulk diffeomorphisms 
are well-known to be gauge transformations in such theories, in the presence 
of boundaries some of these transformations become physical and their
associated charges are nonvanishing.  Characterizing the complete
set of such charges has been the focus of much recent work,
due to various classical and quantum gravitational applications, including 
black hole entropy
\cite{Wald:1993nt, Iyer:1994ys, Strominger1998, 
Carlip_1999, Hawking_2016, Haco:2018ske,Chen:2020nyh,
Chandrasekaran:2020wwn}, 
celestial holography 
\cite{Aneesh:2021uzk, Raclariu:2021zjz, Pasterski:2021rjz}, entanglement entropy
\cite{Donnelly:2014gva, Benedetti:2019uej, David:2022jfd}, and quasilocal 
descriptions of gravitational subregions 
\cite{Donnelly2016a, Speranza2018a, Freidel:2020xyx, Donnelly:2020xgu}.

Due to bulk diffeomorphism invariance, gravitational charges are 
given by integrals over a codimension-2 surface $\partial\Sigma$, located at the 
boundary of a spatial slice $\Sigma$ through the spacetime or subregion
under consideration.  The charges are determined by defining a symplectic
form $\Omega$ for the subregion as the integral of a symplectic current $\omega$ 
over $\Sigma$, and then evaluating $\Omega$ on a diffeomorphism
transformation.  In standard constructions, the charges obtained by this
procedure are of two types.  For diffeomorphisms that map the boundary 
$\partial\Sigma$ into itself, the contraction of this transformation
into the symplectic form yields a total variation, and hence the charges 
satisfy Hamilton's equation and generate a symmetry of the subregion
phase space.  On the other hand, diffeomorphisms with a transverse component
to $\partial\Sigma$ do not yield Hamiltonian charges, since the contraction
of these transformations into the symplectic form is generically not a 
total variation.  
In order to determine a diffeomorphism charge in this case, one 
must split the resulting contraction into a total variation and 
flux term representing the obstruction to integrability of Hamilton's equation.  
This splitting underlies the Wald-Zoupas procedure for determining localized
gravitational charges, and suffers from ambiguities in how to determine
a preferred form of the flux.  
Various proposals for fixing this ambiguity have been considered 
recently, and have included covariance requirements
\cite{Flanagan:2019vbl}, stationarity conditions
\cite{Wald:1999wa,CFP},
and most comprehensively appeals to the variational principle for the subregion
and the associated boundary conditions 
\cite{Chandrasekaran:2020wwn, Chandrasekaran2021long, Bart:2019pno}. 
After fixing this ambiguity, one is further faced with the issue of determining 
the algebra satisfied by these localized charges.  Often this aglebra is defined 
via the Barnich-Troessaert bracket \cite{Barnich:2011mi}, which has been argued to 
coincide with the Poisson bracket of the charges as functions on the 
subregion phase space, subject to certain conditions on how one chooses the 
flux \cite{Chandrasekaran2021long}.  

Recently, a novel proposal for constructing diffeomorphism charges 
has been put forward by Ciambelli, Leigh, and Pai (CLP) \cite{Ciambelli2021}, 
which
avoids the complications of the Wald-Zoupas procedure in determining 
a preferred form of the flux.  This is achieved by enlarging the subregion
phase space to include a set of embedding fields $X$ which parameterize the 
spacetime location of the surface $\partial\Sigma$.  Their procedure
closely parallels the extended phase space introduced by Donnelly and 
Freidel (DF) in \cite{Donnelly2016a} and further generalized by the present 
author in \cite{Speranza2018a}, but differs in a crucial way in their 
choice of subregion symplectic form.  In both cases, the introduction
of embedding fields eliminates the obstruction to integrability of Hamilton's equation
when evaluating the symplectic form on a diffeormorphism.  However, in the DF 
construction, the charges vanish identically, making the transformation
pure gauge, while the charges are nonzero in the CLP extended phase space.  
This leads to the intriguing conclusion that the CLP construction is 
able to define Hamiltonian gravitational charges for all diffeomorphisms
acting near the boundary, without resorting to a Wald-Zoupas procedure
for splitting off a flux term in Hamilton's equation.  

The fact that diffeomorphisms are not pure gauge in the CLP construction
points to a fundamental difference between their proposal and the previous
DF construction.   This implies that the CLP proposal amounts to a genuine 
extension of the phase space by new degrees of freedom associated to the 
embedding fields, while this is not the case for the DF propsal.  
To clarify this point, in section \ref{sec:unitarity} we show that because 
all diffeomorphisms are gauge in the DF construction, there always exists 
a choice of unitary gauge in which the embedding fields are 
eliminated from the phase space description.  In this gauge, the DF extended 
phase space reduces to the standard phase space constructed only from the 
dynamical fields.  On the other hand, the CLP symplectic form
differs from the DF choice by a Jacobson-Kang-Myers (JKM) ambiguity term 
in the Iyer-Wald construction \cite{Jacobson:1993vj, Iyer:1994ys}, 
and this ambiguity does not preserve degeneracy with respect to diffeomorphisms
acting near the boundary.  Because these diffeomorphisms are no longer 
gauge, one cannot access the unitary gauge condition  
through a pure gauge transformation.  Hence, the embedding fields 
cannot be eliminated in the CLP extended phase space, implying that they represent
new physical degrees of freedom.  

This raises an important question about whether the 
CLP charges are defined unambiguously.  Because CLP differs from the DF construction
by an ambiguity term, it is important to determine whether the charges can be 
shifted by further choices of JKM ambiguities.  This question is addressed 
in section \ref{sec:ambig}, where it is demonstrated that there is 
considerable freedom to shift the charges by ambiguities, necessitating a 
further principle for fixing the form of the charges.  However, such ambiguities
also appear in standard covariant phase space constructions without embedding 
fields, and recent developments have shown that these can be resolved 
by appealing to the variational principle for the subregion 
\cite{Compere:2008us,Andrade:2015gja, Andrade:2015fna,
Harlow:2019yfa, Chandrasekaran:2020wwn, Chandrasekaran2021long}.  
We further demonstrate that this resolution carries over to the extended 
phase space, and we derive corrected expressions for the gravitational
charges that match those obtained in recent works on boundaries in the covariant
phase space
\cite{Harlow:2019yfa, Freidel:2020xyx, Chandrasekaran:2020wwn,
Chandrasekaran2021long}.  
This resolution requires that the surface $\partial\Sigma$ be realized
as a cut of a bounding hypersurface $\ns$ for the subregion, and we leave open the 
question as to whether this dependence on a choice of $\ns$ can further 
be eliminated.  Unlike the original CLP construction, once correction terms resolving 
the ambiguities are included in the definition of the phase space, new obstructions
to the integrability of Hamilton's equation can arise.  This somewhat diminishes
the advantage of the CLP construction, but we argue that in some cases 
the obstruction is expected to vanish, and  even when it is nonvanishing,
the Barnich-Troessaert bracket of the charges faithfully reproduces the bracket of the 
diffeomorphism-generating vector fields.  

A final question addressed in section 
\ref{sec:WZ} relates to the reason
behind integrability of Hamilton's equation in the CLP construction.  
In phase space constructions without embedding fields, nonintegrability of 
Hamilton's equation has a simple interpretation in terms of the loss 
of symplectic flux during evolution along the subregion boundary.  
This argument no longer holds in extended phase space constructions since
the effective location of the surface $\partial\Sigma$ does not change relative to 
the dynamical fields, due to the action of diffeomorphisms
on the embedding fields and hence the target location of $\partial\Sigma$.  
To highlight the difference, we further show that a different transformation
can be defined on the CLP extended phase space that is the appropriate analog
of the diffeomorphisms on the non-extended phase space.  It involves a combination
of a diffeomorphism and a change in the embedding map that together fix
the target surface $\partial\Sigma$.  This transformation is shown 
to satisfy the modification of Hamilton's equation that appears 
in the Wald-Zoupas construction, suitably generalized to include contributions
from the embedding fields.

\subsection{Notation for field space}
The construction of phase spaces in this work will utilize 
concepts related to the differential geometry of a field configuration
space, and we briefly review the notation used for calculations performed 
in this space, which largely follows that of reference 
\cite{Chandrasekaran2021long}.
The field configuration space $\fs$ is parameterized by 
all possible configurations of the dynamical fields $\phi$ on the spacetime
manifold $\stm$.  Hence $\fs$ can be viewed as an infinite-dimensional manifold,
on which we can define tensorial objects such as vector fields and 
differential forms.  The variations of the dynamical fields 
at each point $x\in\stm$  define
a basis of one-forms $\delta\phi(x)$ on $\fs$, with the operator $\delta$ playing
 the role of the exterior derivative.  Hence, $\delta$ acting on objects involving 
 one variation will always be taken to be the exterior derivative operator,
 and  implicitly involves an antisymmetrized set of independent variations.  
 We will consistently employ the shorthand $\delta\phi$ for $\delta\phi(x)$.
 A specific linearized variation $\Phi(x)$ of the dynamical field $\phi$ 
 defines a vector field $\h\Phi$ 
on $\fs$, and we will denote the contraction operatation of such a vector field 
into a differential form by $I_{\h\Phi}$, so that in particular, 
$I_{\h\Phi} \delta\phi = \Phi(x)$.  
The field space Lie derivative along a vector field $\h\Phi$ will be 
denoted $L_{\h\xi}$, and when acting on field space differential forms 
satisfies Cartan's magic formula $L_{\h\Phi} = I_{\h\Phi} \delta + \delta I_{\h\Phi}$.

An important set of linearized variations are those corresponding to 
infinitesimal diffeomorphisms.  These will be denoted $\h\xi$ where $\xi^a$ 
is a spacetime vector field, and they satisfy $I_{\h\xi}\delta \phi = \lie_\xi \phi$,
where $\lie_\xi$ is the Lie derivative.  The field space Lie bracket of 
two such vector fields $\h\xi$, $\h\zeta$, constructed from generically 
field-dependent vectors $\xi^a$, $\zeta^a$, is given by 
\cite{Gomes:2018dxs, Chandrasekaran2021long}
\beq\label{eqn:brmod}
[\h\xi,\h\zeta] = -\wh{\brmod{\xi}{\zeta}},
\qquad \brmod{\xi}{\zeta}^a = [\xi,\zeta]^a
-I_\h\xi \delta\zeta^a + I_{\h\zeta}\delta\xi^a.
\eeq
As we will be dealing with diffeomorphism-covariant theories, it is useful
to define an operator $\Delta_\h\xi$ that measures the failure of a 
field-space differential form to transform covariantly.  This is defined to be
\cite{Hopfmuller2018, Chandrasekaran:2020wwn}
\beq\label{eqn:Delxi}
\Delta_\h\xi = L_{\h\xi} -\lie_\xi -I_{\wh{\delta\xi}}.
\eeq

\section{Embedding fields and the extended phase space}

The construction of diffeomorphism charges in this work utilizes an 
extended phase space in which embedding fields $X$ are included as additional
 degrees of freedom in the theory.  Their inclusion into covariant phase space
constructions was initially proposed by Donnelly and Freidel in \cite{Donnelly2016a}
in the case of vacuum general relativity with zero cosmological constant,
and the generalization to arbitrary diffeomorphism-invariant
theories was subsequently derived by the present author in \cite{Speranza2018a}.
Here, we briefly review the standard covariant phase space construction
\cite{Witten:1986qs, Crnkovic1987, Crnkovic:1987tz, Ashtekar1991, LeeWald1990, Wald:1993nt, Iyer:1994ys}, as well
as the extended construction involving the inclusion of embedding fields
\cite{Donnelly2016a, Speranza2018a}.  

The main input for the covariant phase space is the Lagrangian $L[\phi]$, taken to be 
a spacetime differential form of maximal degree constructed from the dynamical fields,
collectively denoted $\phi$. 
These fields will be taken to be tensor fields on spacetime, and can consist of the 
metric $g_{ab}$ and any additional matter fields. 
Varying the Lagrangian with respect to the dynamical fields determines
the equations of motion $E[\phi]$ and symplectic potential current 
$\theta[\phi;\delta\phi]$ 
according to
\beq \label{eqn:delL}
\delta L = E\cdot \delta \phi + d\theta.
\eeq
Taking a second variation of $\theta$ yields the symplectic current $\omega = \delta\theta$,
and its integral over a Cauchy surface $\Sigma$ defines the symplectic form for the 
theory,
\beq \label{eqn:Omega}
\Omega = \int_\Sigma \omega.
\eeq
The current $\omega$ is conserved on-shell, $d\omega = -\delta E\cdot\delta\phi$,
and the covariant phase space is defined 
on the subspace of $\fs$ of all solutions to the equations of motion, 
on which this conservation law holds.

 Under a diffeomorphism $Y:\stm\ra\stm$,
the dynamical fields transform via pullbacks $\phi\ra Y^*\phi$.  
Diffeomorphism-invariance of the theory implies then that the Lagrangian transforms 
covariantly under this transformation,\footnote{With slight modifications,
the formalism can also handle theories whose Lagrangian is only covariant up
to boundary terms \cite{Freidel:2021cjp, Chandrasekaran2021long}.}
\beq \label{eqn:Lcov}
L[Y^*\phi] = Y^*L[\phi].
\eeq
This is simply the statement that the Lagrangian does not depend on any nondynamical
background fields.  This covariance property of the Lagrangian allows one to 
derive a Noether current
\beq 
J_\xi = I_\h\xi \theta - i_\xi L
\eeq
that is conserved on-shell, $d J_\xi = 0$.  Furthermore, since this equation holds for
all vector fields $\xi^a$, one can show that $J_\xi$ can be expressed in terms of 
a potential $Q_\xi$ by the equation 
\cite{W-closed, Iyer:1994ys, Iyer:1995kg}
\beq
J_\xi = dQ_\xi + C_\xi,
\eeq
where $C_\xi = 0$ are combinations of the equations of motion that define the constraints 
of the theory.  Hence, the Noether current $J_\xi$ is exact on-shell.  

The gauge symmetries of the theory can be discerned by examining the degenerate directions
of the symplectic form (\ref{eqn:Omega}).  These correspond to diffeomorphisms acting
in the interior of $\Sigma$.  This is seen by employing the standard on-shell Iyer-Wald 
identity $I_\h\xi \omega = -d(\delta Q_\xi-Q_{\delta\xi} -i_\xi \theta)$, which implies
\beq \label{eqn:IxiOm}
-I_\h\xi\Omega = \int_{\partial\Sigma} (\delta Q_\xi - Q_{\delta\xi} - i_\xi \theta).
\eeq
Since this contraction localizes to a boundary integral, it is immediately apparent
that all diffeomorphisms except for those with support near $\partial\Sigma$ define
gauge transformations of the symplectic form.  On the other hand, diffeomorphisms
acting at $\partial\Sigma$ define physical transformations of the subregion phase space.  
This suggests that the presence of a boundary for the Cauchy surface has promoted
some pure gauge transformations to physical degrees of freedom
\cite{Carlip:1995cd}.

In order to explicitly parameterize these new boundary degrees of freedom, 
Donnelly and Freidel proposed an extension of the dynamical fields of vacuum 
general relativity to include a set of embedding fields $X$
\cite{Donnelly2016a}.  These fields describe 
the embedding of a neighborhood of $\Sigma$ into spacetime, and were included in
the theory to enforce that the symplectic form be fully diffeomorphism-invariant.  
Their construction was reformulated and extended to arbitrary
diffeomorphism-invariant theories in \cite{Speranza2018a} 
by employing the Iyer-Wald formalism 
\cite{Iyer:1994ys}.  
As explained in 
\cite{Donnelly2016a, Speranza2018a, Speranza:2019hkr}, the embedding
fields can be viewed as a diffeomorphism $X:\stmr\ra\stm$ 
from a reference spacetime $\stmr$ into 
$\stm$.  The embedding fields are coupled to the theory only through the pullbacks of the 
dynamical fields $X^*\phi$, in order to preserve covariance of the theory under 
diffeomorphisms.  The variation of any pulled back field satisfies the 
identity \cite{Donnelly2016a, Speranza2018a, Speranza:2019hkr}
\beq\label{eqn:delXphi}
\delta X^*\phi = X^*(\delta\phi + \lie_\rchi \phi)
\eeq
where $\rchi^a$ is a spacetime vector field constructed from the variation of the 
embedding map, and hence is a one-form on field space. Its variation satisfies the 
equation
\beq\label{eqn:flatconn}
\delta \rchi^a + \frac12[\rchi,\rchi]^a = 0,
\eeq
giving it the interpretation of a flat connection on field space, 
viewed as a $\text{Diff}(\stm)$ fiber bundle \cite{Gomes:2016mwl}.
Under a diffeomorphism $Y$, the embedding map transforms to its pullback under $Y^{-1}$,
$X\ra (Y^{-1}\circ X)$.  This transformation law ensures that the pulled back fields 
$X^*\phi$ are fully invariant under diffeomorphisms, since 
$X^*\phi \ra (Y^{-1}\circ X)^* Y^*\phi = X^*(Y^{-1})^* Y^* \phi = X^*\phi$.  
Infinitesimally, this implies from equation (\ref{eqn:delXphi}) that 
\beq
I_\h\xi \rchi^a = 
-\xi^a,
\eeq
since $0=I_\h\xi \delta X^*\phi = X^*(\lie_\xi\phi + \lie_{\left(
I_\h\xi \rchi\right)} \phi)$.

Diffeomorphism invariance of the pulled back
fields then yields a straightforward prescription to couple the embedding fields to theory
\cite{Speranza2018a}: simply write the Lagrangian in terms of the pulled back fields
$X^*\phi$, and full invariance under diffeomorphisms is guaranteed.  The variation
of the extended Lagrangian $L[X^*\phi]$ can then be expressed as 
\beq\label{eqn:delX*phi}
\delta L[X^*\phi] = E[X^*\phi] \cdot \delta X^*\phi + d\theta[X^*\phi;\delta X^*\phi],
\eeq
which shows that $\theta[X^*\phi;\delta X^*\phi]$ serves as a symplectic potential
in the extended phase space.   Using equation (\ref{eqn:delXphi}) and the on-shell
identity $J_\xi = dQ_\xi$, one can show that the extended symplectic potential can
be expressed as 
\beq \label{eqn:thX}
\theta_X\equiv \theta[X^*\phi;\delta X^*\phi] = X^*(\theta +i_\rchi L +dQ_\rchi),
\eeq
where the quantities on the right hand side are functionals of $\phi, \delta\phi$.
The left hand side of this expression is both manifestly invariant under diffeomorphisms
and horizontal  on field space, $I_\h\xi \theta[X^*\phi;\delta X^*\phi]=0$, and hence the 
symplectic form derived from it will share these properties.  
The resulting symplectic form for the extended phase space derived in \cite{Speranza2018a}
is given by
\beq \label{eqn:OmX}
\Omega_X = \int_{X^*\Sigma} \delta \theta_X = \int_{\Sigma}\omega
+ \int_{\partial \Sigma} \left(i_\rchi \theta +\frac12i_\rchi i_\rchi L +
\delta Q_\rchi +\lie_\rchi Q_\rchi\right),
\eeq
and can be shown to reduce to the expression 
derived by Donnelly and Freidel \cite{Donnelly2016a} when
specializing to vacuum general relativity with $\Lambda=0$.  Note that the integrals
over $\Sigma$ and $\partial\Sigma$ depend on the embedding field to determine their
location in spacetime, so that $\int_\Sigma \alpha = \int_{X^*\Sigma} X^*\alpha$.
When computing variations of such integrals, we will always hold the source 
location $X^*\Sigma$ or $X^*\partial\Sigma$ of the embedding map fixed, and hence the variation of any 
such integral will always be understood to including a variation of the embedding map:
\beq
\delta\int_{\Sigma} \alpha =\int_{X^*\Sigma}\delta X^*\alpha = \int_{\Sigma}\left(\delta \alpha + \lie_\rchi \alpha\right)
\eeq
applying (\ref{eqn:delXphi}).

The extended symplectic form $\Omega_X$ annihilates all diffeomorphisms, $I_\h\xi\Omega_X=0$, and hence on this phase space all charges associated with
diffeomorphisms are trivial.  The nontrivial charges that arise in this 
construction are associated with diffeomporphisms of the reference 
space $Z:\stmr\ra\stmr$, which are defined to leave the dynamical fields $\phi$ invariant.  
These transformations act on the embedding fields via $X\ra X\circ Z$, and one can 
show that the infinitesimal transformation corresponding to a vector field 
$w^a$ on $\stmr$ acts via \cite{Donnelly2016a, Speranza2018a}
\beq \label{eqn:Iw}
I_\h{w} \delta \phi = 0, \qquad I_\h{w}\rchi^a = (X_*w)^a\equiv W^a, \qquad I_\h{w}\delta
X^*\phi = \lie_w X^*\phi.
\eeq
Such transformations induce a change in the target location of the embedding map while 
holding fixed the dynamical fields, and hence can be viewed as physical evolution
within a fixed spacetime.  Contracting this transformation into the extended 
symplectic form yields
\beq\label{eqn:IwOmX}
-I_\h{w}\Omega_X = \delta \int_{\partial\Sigma}Q_W -\int_{\partial\Sigma}
\left(Q_{\delta W+[\rchi,W]} - i_W(\theta+i_\rchi L +dQ_\rchi)\right)
\eeq
Here, for transformations $W^a$ that are purely tangential
at $\partial\Sigma$ and which are field-independent
in the sense
$\delta W^a + [\rchi,W]^a = (X_*\delta w) = 0$ will satisfy Hamilton's equation with 
Hamiltonian $H_{\h{w}} = \int_{\partial\Sigma} Q_W$.  For more general transformations,
such as those which deform the surface $\partial\Sigma$ in a 
transverse direction, the remaining terms 
in (\ref{eqn:IwOmX}) prevent the Hamilton's equations from being satisfied for the 
transformation $I_\h{w}$.  These obstruction terms are similar to those appearing 
in equation (\ref{eqn:IxiOm}) for diffeomorphism charges in the non-extended phase space.

\section{Unitary gauge for embedding fields}
\label{sec:unitarity}

The similarity between the charges (\ref{eqn:IwOmX}) in the extended phase space 
and the diffeomorphism charges (\ref{eqn:IxiOm}) is not a coincidence; rather,
it arises due to an equivalence between the two descriptions of the subregion
phase space.  This equivalence is due to the full diffeomorphism-invariance 
of the extended symplectic form $\Omega_X$, and it can be shown that 
the non-extended phase space is simply a gauge fixing of the extended one.  Hence,
the extended phase space construction of Donnelly and Freidel
is fully equivalent to one in which no additional embedding fields are introduced.  

The reason for this equivalence lies in the fact that 
the introduction of embedding fields 
is essentially the Stueckelberg trick for diffeomorphisms
(see, e.g., \cite{Arkani-Hamed2002}).  Equation 
(\ref{eqn:IxiOm}) indicates that the presence of a boundary has 
caused some diffeomorphisms to become physical.  The embedding fields 
are introduced to restore diffeomorphism invariance, just as Stueckelberg
fields can be used to restore gauge invariance in theories with 
massive vector bosons.  However, there always exists a choice of unitary
gauge, in which the embedding field is set to a fixed value $X = X_0$.  
In this gauge, variations of $X$ are also set to zero, which further implies 
$\rchi^a = 0$.  Since all the boundary terms in 
the extended symplectic form (\ref{eqn:OmX}) depend on $\rchi^a$, we see that in
this gauge these boundary contributions drop out, and the symplectic form
reduces to the standard expression in the absence of embedding
fields (\ref{eqn:Omega}).

Note that the $\h{w}$ transformation defined by (\ref{eqn:Iw}) is not
consistent with the unitary gauge condition $\rchi^a=0$.  However,
we are free to redefine this transformation to include an arbitrary
diffeomorphism, since these are pure gauge in the extended phase space.
Choosing this diffeomorphism to be given by $\xi_w = X_* w$, we see that 
the new transformation 
\beq \label{eqn:w0}
\h w_0 = \h w + \h\xi_w
\eeq
satisfies 
\beq\label{eqn:Iw0chi}
I_\h{w_0} \rchi^a = (I_\h{w} + I_{\h{\xi}_w})\rchi^a = (X_* w)^a - (X_* w)^a = 0,
\eeq
consistent with the unitary gauge condition.  Hence, $\h w_0$ is the appropriate
transformation from which to obtain the physical charges in unitary
gauge, and it is straightforward to verify the contraction of this 
transformation into the symplectic form simply reproduces
(\ref{eqn:IxiOm}).  This demonstrates that upon restricting 
to unitary gauge, the physical charges 
constructed in the extended phase space reduce to ordinary diffeomorphism
charges in the non-extended phase space.

This equivalence between the extended phase space in unitary gauge and the 
standard, non-extended phase space raises the question as to whether the introduction
of embedding fields is necessary.  In a recent work
\cite{Ciambelli2021}, Ciambelli, Leigh, and Pai
exhibited a modified construction of an extended phase space in which diffeomorphisms
acting near the boundary are not pure gauge.  This then implies that the unitary gauge 
condition in which the embedding map is fixed to a constant is not accessible via
pure gauge transformations, invalidating the argument for the equivalence between 
the extended and non-extended phase spaces.  The construction 
of CLP utilizes an ambiguity in the Iyer-Wald formalism 
\cite{Jacobson:1993vj, Iyer:1994ys}
to define a modified symplectic
potential whose corresponding symplectic form no longer treats 
boundary diffeomorphisms as 
gauge transformations.  This ambiguity comes from the freedom to shift $\theta$ by 
exact terms $\theta\ra\theta + d\nu$, where $\nu$ is a phase space one-form.  
Their proposal for the extended symplectic form is 
\beq
\theta_{\clp} = X^*(\theta +i_\rchi L),
\eeq
which differs from the previous symplectic potential (\ref{eqn:thX}) by 
the term $dX^*Q_\rchi$,\footnote{The existence of this modified symplectic
potential was mentioned in \cite{Speranza2018a}, but was dismissed on the 
grounds of not admitting boundary diffeomorphisms as degeneracy 
directions.  A crucial insight of the CLP construction is that degeneracy
need not be imposed, and relaxing this condition leads to
useful results.  }
\beq \label{eqn:thCLPthX}
\theta_{\clp} = \theta_X -dX^*Q_\rchi.
\eeq

The ambiguity term $dQ_\rchi$ depends explicitly on $\rchi^a$ without a 
corresponding variation of a dynamical field, unlike the variations appearing
in $\theta_X$ in which all instances of $\rchi^a$ arise from variations
of pulled back fields, $\delta X^*\phi = X^*(\delta \phi +\lie_\rchi \phi)$.
This explicit dependence on $\rchi^a$ is responsible for breaking the full
diffeomorphism-invariance of the symplectic form, ultimately leading to the 
conclusion that the embedding fields cannot be eliminated by gauge-fixing
this phase space.  Hence, unlike the phase space constructed from 
$\theta_X$, the embedding fields represent genuinely new degrees of freedom
in the extended phase space of CLP. 

The main result of \cite{Ciambelli2021} 
is that the phase space constructed from this 
modified symplectic potential yields nonzero charges satisfying Hamilton's 
equation for all diffeomorphisms acting near the boundary, including
those which do not preserve the location of the surface $\partial\Sigma$.
This is in contrast to standard phase space constructions, 
in which there is an obstruction to integrating Hamilton's equation
for such surface-deforming diffeomorphisms, as is evident in 
equations (\ref{eqn:IwOmX}) and (\ref{eqn:IxiOm}). To see how this 
arises, we first note that (\ref{eqn:thCLPthX})
implies that the symplectic form constructed from
$\theta_{\clp}$ can be written as 
\beq
\Omega_{\clp} = \Omega_X -\delta \int_{\partial\Sigma} Q_\rchi
\eeq
with $\Omega_X$ given in equation (\ref{eqn:OmX}).  It was previously
demonstrated that diffeomorphisms are degenerate directions of 
$\Omega_X$, which immediately implies
\beq\label{eqn:IxiOmCLP}
-I_\h\xi \Omega_{\clp} = I_\h\xi \delta \int_{\partial\Sigma}Q_\rchi.
\eeq
To proceed, we note the following general identity satisfied by a pulled-back
phase space form $X^*\alpha$,
\begin{align}
I_\h\xi \delta X^*\alpha &= X^*\left(I_\h\xi \delta \alpha -\lie_\xi\alpha
-\lie_\rchi I_\h\xi \alpha\right) \\
&= X^*\left(L_\h\xi \alpha -\delta I_\h\xi\alpha -\lie_\xi \alpha -\lie_\rchi I_\h\xi \alpha\right) \\
&= -\delta X^* I_\h\xi \alpha + X^*\left(\Delta_\h\xi \alpha +I_\wh{\delta\xi} \alpha\right)
\label{eqn:IxidelX*a}
\end{align}
where the anomaly operator $\Delta_{\h\xi}$ was defined in (\ref{eqn:Delxi}).  
Taking 
$\alpha = Q_\rchi$, we note that because the Noether potential $Q_\xi$ is 
covariantly constructed from the dynamical fields and the vector $\xi^a$, and because 
$\rchi^a$ is a covariant one-form on phase space,\footnote{Covariance 
of $\rchi^a$ follows from the definition of $\Delta_\h\xi$ acting on it,
which gives
\beq \label{eqn:covchi}
\Delta_\h\xi \rchi^a = I_\h\xi \delta \rchi^a + \delta I_\h\xi \rchi^a - \lie_\xi \rchi^a
-I_{\wh{\delta\xi} } \rchi^a
= -\frac12 I_\h\xi [\rchi,\rchi]^a - \delta \xi^a -[\xi, \rchi]^a +\delta \xi^a = 0,
\eeq
applying equation (\ref{eqn:flatconn}) for the variation $\delta\rchi^a$.  
Although not necessary for this work, the fact 
that $\rchi^a$ can be interpreted as a flat $\text{Diff}(\stm)$ 
connection on field space \cite{Gomes:2016mwl}
suggests that one could consider a generalization
in which the connection has curvature, in which case (\ref{eqn:flatconn})
would be modified to $\delta\rchi^a + \frac12[\rchi,\rchi]^a = \rho^a$, where $\rho^a$ 
is a field-space two form defining the curvature of the connection.  The proof 
of covariance continues to go through \cite{Freidel:2021dxw}, 
since (\ref{eqn:covchi}) would be corrected 
by a term $I_\h\xi \rho^a$, which vanishes since the curvature of a connection
is always a horizontal form \cite{Choquet2004}.} 
we have that $\Delta_\h\xi Q_\rchi=0$, and so for field-independent diffeomorphisms
($\delta\xi^a=0$),
equation (\ref{eqn:IxiOmCLP}) reduces to
\beq \label{eqn:CLPHam}
-I_\h\xi \Omega_\clp = \delta \int_{\partial\Sigma} Q_\xi,
\eeq
showing that this transformation satisfies Hamilton's equation with a Hamiltonian
charge
\beq
H_\xi = \int_{\partial\Sigma} Q_\xi.
\eeq

The introduction of embedding fields therefore has been shown to yield genuine 
Hamiltonian charges associated with all diffeomorphisms.  This is to be contrasted 
with constructions that involve no embedding fields, in which case diffeomorphisms
which move the surface are associated with Wald-Zoupas charges, which satisfy
a modification of Hamilton's equations that allows for additional flux contributions
\cite{Wald:1999wa, Chandrasekaran:2020wwn, Chandrasekaran2021long}.  
Note that it was already apparent in the 
extended phase space of Donnelly and Freidel that embedding fields yield Hamiltonian
diffeomorphism 
charges; however, in their construction, all of these charges vanish identically.
The important novelty of the CLP construction is their modification of the phase space
symplectic form to arrive at nonzero charges while 
simultaneously preserving  the Hamiltonian property.  

Since these charges satisfy Hamilton's equation, the Poisson bracket between two
of them can be computed by contracting the corresponding vector fields into the 
symplectic form.  One verifies straightforwardly that
\beq
\{ H_\xi, H_\zeta\} =  - I_\h\xi \delta H_\zeta
 = -\int_{\partial\Sigma} (I_\h\xi \delta Q_\zeta - \lie_\xi Q_\zeta)
 = -\int_{\partial\Sigma} \Delta_\h\xi Q_\zeta
 = \int_{\partial\Sigma} Q_{[\xi,\zeta]}
\eeq
where the last equality uses that the only noncovariance in $Q_\zeta$ comes from
its dependence on the fixed vector $\zeta^a$, so $\Delta_\h\xi Q_\zeta 
= Q_{\Delta_\h\xi \zeta} = -Q_{[\xi,\zeta] -I_\h\xi\delta \zeta}$,
and then noting that $\delta\zeta^a =0$.  This reproduces the result of 
\cite{Ciambelli2021, Ciambelli:2021vnn}
that Poisson brackets of the charges yield a representation
of the vector field algebra under Lie brackets.  

The nonzero charges correspond to diffeomorphisms tangential
to the surface $\partial\Sigma$, 
pointwise $\text{SL}(2,\mathbb{R})$ transformations in the normal plane, 
and two
independent surface deformations that move the surface in the transverse directions.
This algebra has the structure of the Lie algebra of the group
$\text{Diff}(\partial\Sigma)\ltimes(\text{SL}(2,\mathbb{R})
\ltimes \mathbb{R}^2)^{\partial\Sigma}$, and was first identified in 
\cite{Speranza2018a}, and subsequently explored and expanded upon
in \cite{Ciambelli:2021vnn, Freidel:2021cjp}.  Furthermore, this algebra 
appears universally in any diffeomorphism-invariant theory \cite{Speranza2018a},
provided certain choices are made to resolve the ambiguities in the Iyer-Wald formalism.
Resolving these ambiguities leads to important corrections to the charges, 
and hence we next turn to understanding how this resolution can be applied in the 
new proposal of CLP.

\section{Ambiguities and their resolution}
\label{sec:ambig}

Ambiguities arise in the Iyer-Wald formulation of the covariant phase space due to 
certain quantities being defined only up to the addition of spacetime exact forms.  
Specifically, the Lagrangian $L$ and symplectic potential $\theta$ can both be shifted
according to
\begin{align}
L&\ra L + d a \label{eqn:Lamb}\\
\theta &\ra \theta + \delta a + d\nu \label{eqn:thamb}
\end{align}
without affecting the relation (\ref{eqn:delL}).  Resolving these ambiguities is 
of crucial importance when constructing charges, since, for example, the difference between
the DF and CLP extended phase spaces is given by such an ambiguity.  
To further emphasize the issue presented by these ambiguities, note that the existence 
of a covariant spacetime vector field $\rchi^a$ constructed from variations of the 
embedding fields allows for the construction of a wide variety of covariant two-forms
$\Amb_\rchi$ that can be used as ambiguities to change the extended phase space symplectic
form.  The new symplectic potential would then be $\theta_A = \theta_{\clp} + dX^*A_\rchi$,
resulting in a new extended symplectic form
\beq
\Omega_A = \Omega_{\clp} + \delta\int_{\partial\Sigma} A_\rchi.
\eeq
Covariance implies that $\Delta_\h\xi A_\rchi = 0$, and hence the derivation of 
section 
\ref{sec:unitarity} can be repeated to derive a shifted diffeomorphism
charge 
\begin{align}
-I_\h\xi \Omega_A &= \delta H_\xi^A \\
H_\xi^A &= \int_{\partial\Sigma} (Q_\xi -A_\xi)
\end{align}
Additionally, any ambiguity terms $\nu$ constructed solely from the dynamical fields
will also shift the charges.  We  further allow for the ambiguity
terms to be generically noncovariant, but as demonstrated below,
this tends to spoil the integrability of Hamilton's equation.

Recently, it has been understood that these ambiguities can be resolved by appealing to
the variational principle for a subregion in spacetime 
\cite{Compere:2008us,Andrade:2015gja, Andrade:2015fna,
Harlow:2019yfa, Chandrasekaran:2020wwn, Chandrasekaran2021long}.  
This resolution requires
$\partial\Sigma$ to arise as a cut of a hypersurface $\ns$ bounding the 
spacetime subregion
$\sr$ under consideration.  Given the additional structure
provided by the hypersurface $\ns$, one looks for a decomposition of the pullback of $\theta$
to $\ns$, denoted $\p \theta$, 
of the form 
\beq\label{eqn:thdecomp}
\p \theta = -\delta \ell + d\beta +\beom.
\eeq
The flux term $\beom$ is the quantity that would be set to zero by boundary conditions
in the variational principle for the subregion.
Note that the identification of the boundary condition is used to 
single out a preferred form of $\beom$, but 
we do not assume such boundary conditions have been imposed, since 
they restrict the dynamics in finite subregions.  
To obtain an unambiguous decomposition, criteria must be 
given for fixing the 
form of the flux.  For example, in general relativity, given a subregion
bounded by a timelike surface, we can require the flux be in Dirichlet form
$\beom = \pi^{ij}\delta h_{ij}$.  The term $\ell$ appearing in (\ref{eqn:thdecomp})
is then used as the boundary term when constructing the subregion action,
\beq
S = \int_\sr L + \int_\ns \ell + \ldots
\eeq
where the dots denote additional terms at past, future, or higher
codimension boundary components.  The symplectic form for the subregion
also receives a correction from the quantity $\beta$ in (\ref{eqn:thdecomp}), 
\beq
\Omega = \int_\Sigma \omega -\delta\int_{\partial\Sigma} \beta.
\eeq
One can demonstrate that this symplectic form is invariant under the ambiguities
described in equations (\ref{eqn:Lamb}) and (\ref{eqn:thamb}), and leads to 
ambiguity-free expressions for the charges as well
\cite{Harlow:2019yfa, Chandrasekaran:2020wwn, Chandrasekaran2021long}.  

This resolution of the ambiguities continues to apply in the context of the 
extended phase space, and we will show that it justifies the procedure
employed by CLP in their modification of the subregion symplectic form.  Furthermore,
we will find that specific corrections to the charges appear, which match the 
corrections to charges explored in several recent works 
\cite{Harlow:2019yfa, Freidel:2020xyx, Chandrasekaran:2020wwn, 
Freidel:2021cjp, Chandrasekaran2021long}.  
Beginning with the extended Lagrangian $L[X^*\phi] = X^* L[\phi]$, we can determine the
extended symplectic potential from the variation
\beq
\delta X^*L = X^*(\delta L+ di_\rchi L) = X^*(E\cdot \delta \phi) + dX^*(\theta + i_\rchi L).
\eeq
Here, $\theta_{\clp} = X^*(\theta + i_\rchi L)$ appears naturally when parametrizing the 
variations in terms of $\delta \phi$ as opposed to $\delta X^*\phi$ appearing in equation
(\ref{eqn:delX*phi}).  
We then proceed to carry out the decomposition (\ref{eqn:thdecomp}) for $\theta_{\clp}$ pulled 
back to the boundary hypersurface $\ns$.  Taking $\ell$, $\beta$, and $\beom$ to be defined 
by the decomposition for $\theta$ alone, we have that
\begin{align}
\p{\theta_\clp} &= X^*\left(-\delta \ell + d\beta +\beom +i_\rchi L\right) \\
&= -\delta\left(X^*\ell\right) + X^*\left(\lie_\rchi \ell + d\beta + \beom +i_\rchi L\right)\\
&= -\delta X^*\ell + dX^*\left(\beta+i_\rchi \ell\right) + X^*\left(\beom +i_\rchi(L+d\ell)\right)
\end{align}
giving the expressions for the boundary, corner, and flux terms in the extended phase space,
\begin{align}
\ell_{\ext} & = X^* \ell \\
\beta_{\ext} &= X^*(\beta + i_\rchi \ell) \label{eqn:betaex}\\
\beom_{\ext} &= X^*(\beom + i_\rchi(L +d\ell)). \label{eqn:beomex}
\end{align}

To argue for the uniqueness of this decomposition, we must examine the boundary
condition
implied by this choice of the flux $\beom_{\ext}$.  The first term $X^* \beom$ is 
simply the pullback of the flux term that appears 
before adding the embedding fields, and hence 
will vanish for the same set of boundary conditions that would make $\beom$ itself
vanish.  The second term $X^* i_\rchi(L+d\ell)$ involves the flux due to variations of the 
embedding field $X$.  Because it depends only on $\rchi^a$ and not its derivatives, this 
term takes a Dirichlet form for the embedding field $X$, since it vanishes if the 
Dirichlet condition $\delta X = 0$ is imposed at the boundary.\footnote{To see
the how the vanishing of $\rchi^a$ is related to the Dirichlet 
condition for the embedding fields, we can take $X^\mu$ to be a set 
of scalar functions defining the embedding map.  Then the coordinate expression
for the pullback $X^*\rchi^a$ is given by \cite{Speranza:2019hkr}
\beq
X^*\rchi^a = \delta X^\mu (X^*\partial_\mu^a).
\eeq
}  We will see presently that taking the Dirichlet form
for the embedding field flux leads to the expected expression for the improved 
charges.  

The corner terms $\beta_{\ext}$ in (\ref{eqn:betaex}) defines a correction to the 
extended symplectic form, which is now given by
\beq \label{eqn:Omex}
\Omega_{\ext} = \Omega_{\clp} -\delta\int_{\partial\Sigma} (\beta+i_\rchi\ell).
\eeq
To determine the diffeomorphism charges, we combine the expression (\ref{eqn:CLPHam})
for the CLP symplectic form with the general relation (\ref{eqn:IxidelX*a}) to 
derive
\begin{align}
-I_\h\xi \Omega_{\ext} &= 
\delta\int_{\partial\Sigma} h_\xi +\int_{\partial\Sigma}\left(
\Delta_\h\xi\left(\beta+i_\rchi \ell \right)
- h_{\delta\xi} \right) \label{eqn:ham+obs} \\
h_\xi &= Q_\xi + i_\xi \ell - I_\h\xi \beta \label{eqn:hxi}
\end{align} 
where we have retained the terms involving $\delta\xi$.  In order for Hamilton's equation
to be satisfied, not only does the diffeomorphism need to be field-independent $\delta\xi = 0$,
the quantity $\Delta_\h\xi(\beta+i_\rchi \ell)$ must also vanish.  The charge
is then given by the integral of the charge density $h_\xi$ (\ref{eqn:hxi}), which
includes corrections coming from $\ell$ and $\beta$.  This corrected charge coincides
with the expression derived by Harlow and Wu in their work on covariant phase space 
with boundaries
\cite{Harlow:2019yfa}\, and has appeared subsequently in a variety of other contexts
\cite{Freidel:2020xyx, Chandrasekaran:2020wwn, 
Freidel:2021cjp, Chandrasekaran2021long}.  

Since the improved charges have nontrivial integrability conditions in order to satisfy
Hamilton's equation, it is interesting to investigate when this condition is satisfied.  
While less is know for generic diffeomorphism-invariant theories, in the case of 
general relativity it has been shown that when 
choosing $\beom$ to be of Dirichlet form, $\Delta_\h\xi \beta = 0$ for timelike and 
null boundaries if the transformation generated by $\xi^a$ preserves the bounding
hypersurface $\ns$.  Such transformations also satisfy $\Delta_\h\xi \ell = 0$ in the timelike
case, but this term can be nonzero when working with null surfaces
\cite{Chandrasekaran:2020wwn}.  In fact, a nonzero
value of $\Delta_\h\xi\ell$ has been shown to be a necessary ingredient for extensions to 
appear in the Poisson bracket algebra of the charges, so it is interesting here to 
see it appearing as an obstruction to integrability of Hamilton's equation within the 
extended phase space.  For transformations with a transverse component to $\ns$, in general
we would expect both $\Delta_\h\xi\beta$ and $\Delta_\h\xi \ell$ to be nonzero.  
Another context in which there may be a nonzero contribution from $\Delta_\h\xi\beta$ is 
in higher curvature theories, where in order to arrive at the universal embedding 
subalgebra 
$\text{Diff}(\partial\Sigma)\ltimes(\text{SL}(2,\mathbb{R})
\ltimes \mathbb{R}^2)^{\partial\Sigma}$ as the only nonzero charges, a specific 
contribution to $\beta$ must be added that is not spacetime covariant
\cite{Speranza2018a}.  It would 
be interesting to investigate the extent to which this obstructs the construction
of unambiguous Hamiltonian charges in the higher curvature context.  

In the case that the obstruction 
\beq \label{eqn:obs}
\flx_\h\xi = -\int_{\partial\Sigma}\left(\Delta_{\h\xi}(\beta + i_\rchi \ell) - h_{\delta\xi}\right)
\eeq
is nonzero, we can 
still consider the quantity 
\beq \label{eqn:Hxi}
H_\xi = \int_{\partial\Sigma} h_\xi
\eeq
as a function on the subregion phase space.  It will not satisfy Hamilton's equation
for the diffeomorphism transformation due to the obstruction term (\ref{eqn:obs}).  
Instead, equation (\ref{eqn:ham+obs}) represents a modification of Hamilton's equation, where 
the obstruction  appears as a flux of the local charge.  In this case, 
one can still seek to compute the Poisson brackets of the charges $H_\xi$, in a similar
manner to the procedure explained in section 3 of 
\cite{Chandrasekaran2021long}.  
Letting $\n A, \n B, \ldots$ denote abstract indices on phase space and choosing
an inverse $\Omega_{\ext}^{\n A\n B}$ for the symplectic form,\footnote{Due
to degeneracies, $\Omega^{\ext}_{\n A\n B}$ is not invertible, but we 
can construct a partial inverse that satisfies
$\Omega_{\ext}^{\n A\n B}\Omega^{\ext}_{\n B\n C}\h\xi^{\n C}$. } this Poisson bracket is computed to be
\begin{align}
\{H_\xi, H_\zeta\} 
&= 
\Omega_{\ext}^{\n A\n B}(\delta H_\xi)_{\n A} (\delta H_\zeta)_{\n B} 
\\
&= 
\Omega_{\ext}^{\n A\n B} 
\left(\Omega^{\ext}_{\n A\n C} \h\xi^{\n C} + (\flx_{\h\xi})_{\n A} \right)
\left(\Omega^{\ext}_{\n B \n D} \h\zeta^{\n D} + (\flx_{\h\zeta})_{\n B}\right) 
\\
&=
\{ H_\xi, H_\zeta\}_{\text{BT}} 
+ \Omega_{\ext}^{\n A\n B}(\flx_{\h\xi})_{\n A}(\flx_{\h\zeta})_{\n B}
\label{eqn:pbflx}
\end{align}
where we have defined the Barnich-Troessaert (BT) bracket 
\cite{Barnich:2011mi}
\beq \label{eqn:BTbrack}
\{ H_\xi, H_\zeta\}_{\text{BT}} = -I_\h\xi \delta H_\zeta + I_{\h\zeta}\flx_\h\xi.
\eeq

The BT bracket will coincide with the Poisson bracket, provided that the 
flux terms in (\ref{eqn:pbflx}) can be shown to drop out.  In 
\cite{Chandrasekaran2021long}, it was argued that this term will vanish if the flux 
term $\beom$ in the decomposition of the symplectic potential is 
in Dirichlet form.  In the present context, if all terms appearing in the 
integrand (\ref{eqn:obs}) for $\flx_\h\xi$ only involve undifferentiated variations
of the dynamical fields and $\rchi^a$, we would similarly expect the term quadratic in 
the flux in (\ref{eqn:pbflx}) to drop out.  For example, if we 
specialize to general relativity, consider a field-independent 
transformation $\delta\xi^a=0$, and 
take the vector field $\xi^a$ to be tangent to the hypersurface $\ns$, the 
only remaining term in the flux will be the integral of $i_\rchi\Delta_\h\xi \ell$.  
The contribution involving $\flx_\h\xi$ in (\ref{eqn:pbflx}) will vanish then
as long as
\beq
\Omega_{\ext}^{\n A \n B}\rchi_{\n A}^a \rchi_{\n B}^b = 0,
\eeq
which holds assuming that the embedding fields commute among 
themselves,\footnote{Intriguingly, there remains a possibility that this commutator not
vanish if one considered a noncommutative geometry setup, as occurs in some approaches
to quantum gravity and string theory (see, e.g.\
\cite{Connes1994, Chamseddine:1996rw, Seiberg:1999vs}), in which case the 
Poisson bracket of localized charges would be corrected from the BT bracket.}
\beq
\{ X^\mu, X^\nu\} = 0.
\eeq

The BT bracket of the charges can be evaluated  directly from equation 
(\ref{eqn:BTbrack}) to obtain a charge representation theorem.  Using 
that $h_\zeta = -I_\h\zeta(Q_\rchi +i_\rchi \ell + \beta)$
and $[\Delta_\h\xi, I_\h\zeta] = -I_{\brmod{\xi}{\zeta}} + I_{\wh{I_\h\zeta\delta\xi}}$
\cite{Chandrasekaran2021long}, and recalling that $\Delta_{\h\xi} Q_{\rchi} = 0$,
we find
\begin{align}
\{H_\xi, H_\zeta\}_\text{BT} &=
-\int_{\partial\Sigma}\left( I_\h\xi\delta h_\zeta -\lie_\xi h_\zeta
+I_\h\zeta\Delta_\h\xi(\beta +i_\rchi \ell) - I_\h\zeta h_{\delta\xi}\right)
\\
&=
-\int_{\partial\Sigma}\left( -\Delta_\h\xi I_\h\zeta (Q_{\rchi}+i_\rchi \ell +\beta) 
+ I_\h\zeta\Delta_\h\xi(\beta+i_\rchi \ell) 
- h_{I_{\h\zeta}\delta\xi} \right)
\\
&= \int_{\partial\Sigma} h_{\brmod{\xi}{\zeta}}  = H_{\brmod{\xi}{\zeta}}
\label{eqn:HxizetaBT}
\end{align}
where the modified Lie bracket $\brmod{\cdot}{\cdot}$ of field-dependent 
vector fields was defined in (\ref{eqn:brmod}).
Hence, the BT bracket of the localized charges reproduces the algebra satisfied by the 
vector fields under the bracket $\brmod{\cdot}{\cdot}$.
Crucially, no extension terms appear in the bracket, unlike the examples of 
localized charges constructed without introducing embedding fields.  For the 
non-extended phase spaces, the BT bracket of the charges instead satisfies the relation
$\{H_\xi, H_\zeta\}_{\text{BT}} = H_{\brmod{\xi}{\zeta}} + K_{\xi,\zeta}$, where the extension
$K_{\xi,\zeta}$ is generically nonzero.  This therefore generalizes the results of 
\cite{Ciambelli2021, Ciambelli:2021vnn}
that the diffeomorphism charges represent the vector
field algebra without extensions to the case of ambiguity-free charges for subregions
with boundary.  

It is worth mentioning a related recent construction in the context 
of a non-extended phase space in which the bracket of 
the charges also represents the vector field bracket without
extension \cite{Freidel:2021cjp}.  This work defines a bracket 
of the charges that subtracts the extension term
$K_{\xi,\zeta}$ from the BT bracket, so that, by
definition, the bracket yields a representation
of the modified vector field bracket $\brmod{\cdot}{\cdot}$
without extension.  While the original construction
of this bracket did not arise from a Poisson bracket on a phase space,
it has been pointed out in \cite{Freidel:2021dxw}
that it does arise in the CLP extended phase
space applied to covariant charges, 
and simply coincides with the Poisson bracket 
of the diffeomorphism charges.  The representation
theorem (\ref{eqn:HxizetaBT}) extends this result to
the improved, ambiguity-free charges, with the caveat that the 
BT bracket computed in the relation may not coincide with the 
Poisson bracket on the subregion phase space unless 
the flux terms in (\ref{eqn:pbflx}) commute.

As a final aside, we mention that an additional ambiguity can arise 
due to the fact that fixing $\beom$ does not uniquely specify
$\ell$ and $\beta$ in the decomposition (\ref{eqn:thdecomp}),
since shifts of the form $\ell\ra\ell+de$, $\beta\ra\beta+\delta e$ do not
affect this equation.  A proposal for resolving this final ambiguity
was made in \cite{Chandrasekaran:2020wwn, Chandrasekaran2021long}, 
in which one must further decompose
$\beta$ as 
\beq
\beta = -\delta c + \vep
\eeq
and give a criterion for determining the corner flux $\vep$.  This then 
leads to a correction to the charges, and this contribution carries over to the 
charges in the extended phase space.  Noting that $[\delta, \Delta_{\h\xi}] = 
\Delta_{\wh{\delta\xi}}$ \cite{Chandrasekaran2021long}, we find that 
\begin{align}
\int_{\partial\Sigma} \Delta_{\h\xi}\beta &=
\int_{\partial\Sigma}\left(-\delta \Delta_{\h\xi}c +\Delta_{\wh{\delta\xi}} c+
\Delta_{\h\xi}\vep\right) \\
&=
-\delta \int_{\partial\Sigma}\Delta_{\h\xi} c+\int_{\partial\Sigma}\left(
\Delta_{\h\xi}\vep + i_\rchi dc +\Delta_{\wh{\delta\xi}} \vep\right)
\end{align}
Then defining the improved charge density 
\beq\label{eqn:htilde}
\tilde h_{\xi} = h_\xi -\Delta_{\h\xi} c,
\eeq
we see that equation (\ref{eqn:ham+obs}) can be re-expressed as 
\beq
-I_{\h\xi}\Omega_{\ext} = \delta \tilde H_{\xi} -\tilde\flx_{\h\xi}
\eeq
with 
\begin{align}
\tilde H_{\xi} &= \int_{\partial\Sigma} \tilde h_\xi \label{eqn:Htilde}\\
\tilde \flx_\xi &= -\int_{\partial\Sigma}\left(
\Delta_{\h\xi}(\vep +i_\rchi(\ell+dc)) -\tilde h_{\delta\xi}\right)
\end{align}
It would be interesting to study these corner improvements 
in more detail, since, for example, they may provide a way of eliminating 
the dependence of the ambiguity-free charges on the choice of the hypersurface
in which $\partial\Sigma$ is embedded.

\section{Wald-Zoupas charges}
\label{sec:WZ}

One aspect of the Hamiltonian charges obtained in the CLP extended phase space 
that is initially surprising is that the charges satisfy Hamilton's equation for 
surface-deforming
diffeomorphisms that move the bounding surface $\partial \Sigma$.
In standard constructions that do not employ embedding fields, such surface 
deformations fail to satisfy Hamilton's equation, and there is a simple physical
explanation why this occurs.  Surface deformations 
describe transformations of the dynamical fields corresponding to 
evolution along the boundary of the subregion, and during this evolution
one generically expects flux of gravitational and matter degrees of freedom 
through the boundary.  This flux appears as an obstruction to integrability
of Hamilton's equation, and hence one does not expect to be able to 
obtain Hamiltonian charges in this case.  Instead, one can employ a Wald-Zoupas 
procedure to isolate a term in this equation that can be identified 
as the localized charge, and the obstruction term is used to construct the flux
which parameterizes the failure of the local charge to be conserved as one 
evolves along the boundary 
\cite{Wald:1999wa, CFP, Chandrasekaran:2020wwn, Chandrasekaran2021long}.  This procedure yields 
well-defined charges as long as a criterion for specifying the flux is given,
and making such a choice is equivalent to determining the decomposition
(\ref{eqn:thdecomp})
of the symplectic potential.   For example, one can appeal to the action
principle for the subregion to fix the form of the flux
\cite{Chandrasekaran:2020wwn, Chandrasekaran2021long}.  

In the extended phase space, this physical argument for nonintegrability of 
Hamilton's equation no longer applies.  The reason is that the location
of the surface $\partial\Sigma$ is now determined as the image of the embedding 
map $X$, 
which implies that a diffeomorphism now changes the location of the 
surface in addition to evolving the dynamical fields.  
Hence, when viewed relationally to the dynamical fields, 
the effective location of the surface 
is fixed, leaving no room for loss of symplectic flux during the diffeomorphism
transformation.  This holds in both the DF extended phase space 
in which diffeomorphisms are pure gauge, as well as in the 
CLP extended phase space containing nonzero diffeomorphism charges.  

To better understand how fixing the effective relational location of the surface
produces integrable Hamiltonian charges, it is helpful to see how one can 
obtain Wald-Zoupas charges in the extended phase space, which instead 
satisfy the modified Hamilton equation involving a flux term.  These charges 
would arise from a transformation on phase space that changes the relational
location of the surface relative to the dynamical fields.  This can be achieved
by changing the embedding map $X$ while holding the dynamical fields fixed.  In 
the standard extended phase space, this corresponds to the 
$\h w$ transformation described in equation (\ref{eqn:Iw}), resulting in the 
desired modified Hamilton's equation (\ref{eqn:IwOmX}).  This equation
can further be modified to include corrections for resolving the ambiguities
in phase space.  

On the other hand, the $\h w$ transformation alone does not yield the appropriate 
relation for the CLP extended phase space, as might be exptected due to the 
difference in symplectic forms.  Instead, the appropriate transformation
is a combination of a spacetime diffeomorphism and a change in the embedding
map which together fix the spacetime location of the target surface $\partial\Sigma$.
This turns out to be none other than the $\h w_0$ transformation defined in
equation (\ref{eqn:w0}) in the unitary gauge description of the
extended phase space.  The condition that it fix the target location
is equivalent to equation (\ref{eqn:Iw0chi}).  

To confirm this is the desired transformation, we evaluate the contraction
\begin{align}
-I_{\h w_0} \Omega_{\ext}
&= -I_{\h\xi_w}\Omega_{\ext} - I_{\h w}\Omega_{\ext}.
\end{align}
The first term is given by equation (\ref{eqn:ham+obs}).  The second term can be 
evaluated by noting that the bulk contribution to $\Omega_{\ext}$ depends only on 
variations of the dynamical fields, which have zero contraction with $\h w$.  
The remaining boundary contributions can be evaluated using that the 
boundary term in $\Omega_{\clp}$ is given by the integral of $i_\rchi \theta
+\frac12i_\rchi i_\rchi L$ \cite{Ciambelli2021}, 
which combines with the remaining terms in the expression
(\ref{eqn:Omex}) for $\Omega_{\ext}$ to give 
\begin{align}
-I_{\h w}\Omega_{\ext} &=
-I_{\h w}\int_{\partial\Sigma}\left(i_\rchi\theta+\frac12i_\rchi i_\rchi L
+\delta \beta + \lie_\rchi\beta -\frac12 i_{[\rchi,\rchi]} \ell
-i_\rchi\delta \ell +\lie_\rchi i_\chi \ell\right) \\
&=
-\int_{\partial\Sigma}\left( i_\xi \theta +i_\xi i_\rchi L
+\lie_\xi \beta -i_{[\xi,\rchi]}\ell-i_\xi\delta\ell
+ \lie_\xi i_\rchi \ell - \lie_\rchi i_\xi \ell\right)
\\
&=
-\int_{\partial\Sigma}\left(i_\xi \beom+ i_\xi i_{\rchi}(L+d\ell)\right),
\label{eqn:IwOmex}
\end{align}
where we employed the shorthand $\xi$ for $\xi_w$, and 
the last line applied the decomposition (\ref{eqn:thdecomp}) of $\p \theta$.  
Combining with (\ref{eqn:ham+obs}), we arrive at the expression for the contraction
of $\h w_0$ into the extended symplectic form,
\begin{align}
-I_{\h w_0}\Omega_{\ext} &= \delta\int_{\partial\Sigma} h_\xi
-\int_{\partial\Sigma}\left(i_\xi \beom
+ i_\xi i_\rchi(L+d\ell) -\Delta_\h\xi(\beta+i_\rchi \ell)
 + h_{\delta\xi}\right). \label{eqn:Iw0Omex}
\end{align}
Note that the combinations $\beom + i_\rchi(L +d\ell)$ and $\beta + i_\rchi \ell$
are the same as those appearing in the flux and corner terms $\beom_{\ext}$ 
(\ref{eqn:beomex}) and 
$\beta_{\ext}$ (\ref{eqn:betaex}) in the decomposition of the extended symplectic form
$\theta_{\clp}$.

Equation (\ref{eqn:Iw0Omex}) is the expected form of the modified Hamilton equation
satisfied by Wald-Zoupas charges.  The first term is the total variation of the 
localized charge, which is precisely the same form as the 
ambiguity-free diffeomorphism
charge (\ref{eqn:Hxi}).  The remaining terms represent the flux and parameterize
the failure of the $\h w_0$ transformation to preserve the subregion symplectic form.
Comparing to the analogous equation for Wald-Zoupas constructions in the non-extended 
phase space \cite{Chandrasekaran:2020wwn, Chandrasekaran2021long}, 
we see that the same terms 
appear in the expression for the flux, up to the additional terms 
$\Delta_\h\xi i_\rchi \ell+i_\xi i_\rchi(L+d\ell)$ constructed from
$\rchi^a$.  Further comparing to equation (\ref{eqn:ham+obs}) satisfied by the 
pure diffeomorphism transformation, we see that the terms 
$\Delta_{\h\xi}(\beta+i_\rchi\ell) - h_{\delta\xi}$ represent 
obstructions to integrability due to the pure diffeomorphism, while the 
remaining terms $-i_\xi(\beom +i_\rchi(L+d\ell))$ are the obstruction due to the 
change in the embedding, as indicated by equation (\ref{eqn:IwOmex}).  

Together, equations (\ref{eqn:ham+obs}) and (\ref{eqn:Iw0Omex}) demonstrate 
that the CLP extended phase space can describe both diffeomorphism charges and 
Wald-Zoupas localized charges by changing the details of the phase space 
transformation.  Both charges are given by the integral of the charge 
density $h_\xi$ defined in (\ref{eqn:hxi}),\footnote{Nevertheless, 
the two types of charges may differ as functions on 
phase space, depending on how the field-dependence of $\xi^a$ is chosen.}
and in general neither type 
of charge satisfies Hamilton's equation due to specific obstruction
terms to integrability.

\section{Discussion}

This work has shown that the gravitational charges constructed in the 
CLP extended phase space can be improved to ambiguity-free expressions
by applying recent results for handling boundaries in the 
covariant phase space.  Although the ambiguity resolution 
generically produces a new obstruction to integrability of Hamilton's equation, 
charges can nevertheless be identified and given an algebra structure through
BT bracket.  This algebra gives a representation of the vector field
field-dependent Lie bracket without extension.  Furthermore, the CLP 
extended phase space is flexible enough to describe  
standard diffeomorphism charges, as well as Wald-Zoupas-like charges
whose obstruction to integrability is precisely analogous to that 
encountered in non-extended phase spaces.  

An immediate question to be addressed in the present construction
is whether the extension of $\partial\Sigma$ to a hypersurface $\ns$ 
is necessary in constructing ambiguity-free charges.  On the one hand, 
this choice is natural when constructing an action principle for an open
subregion in spacetime $\sr$ and taking $\partial\Sigma$ to simply be a cut 
of a component $\ns$ of the boundary.  On the other hand, it seems more 
natural for the charges to be completely covariant with respect to the 
codimension-2 surface $\partial\Sigma$, since such a surface 
determines a subregion of spacetime via the causal development of 
an infilling hypersurface $\Sigma$.  It is also notable that the symmetry group 
$\text{Diff}(\partial\Sigma)
\ltimes (\text{SL}(2,\mathbb{R})\ltimes \mathbb{R}^2)^{\partial\Sigma}$ is 
naturally associated with covariance with respect to the codimension-2 surface
\cite{Ciambelli:2021vnn},
while introducing the bounding hypersurface $\ns$ would be expected to 
break this down to a subgroup.
A possibility for maintaining corner covariance is to use an appropriate 
corner improvement to the charges as described in equations (\ref{eqn:htilde}) and (\ref{eqn:Htilde}),
and references \cite{Chandrasekaran:2020wwn, Chandrasekaran2021long}.  
It would be useful to take up this 
question in more detail.  

Another generalization would be to investigate higher curvature 
theories. The Iyer-Wald formalism employed in this work
immediately applies to any diffeomorphism-invariant theory.  
However, in order to maintain the same universal symmetry 
group $\text{Diff}(\partial\Sigma)
\ltimes (\text{SL}(2,\mathbb{R})\ltimes \mathbb{R}^2)^{\partial\Sigma}$,
specific choices must be made for resolving the ambiguities
\cite{Speranza2018a}.  These resolution terms can introduce 
noncovariance into the charges and obstruct integrability of Hamilton's equation.  
Nevertheless, since the charges and brackets still appear to be well-defined 
in the presence of these obstructions, one still has a consistent construction
of improved gravitational charges.  It would be interesting to  
compute the explicit form of the obstruction to integrability in order to 
to better understand how the CLP construction extends to 
other diffeomorphism-invariant theories.

At a practical level, the CLP charges provide a possible way to better
understand charges constructed at asymptotic boundaries on which leaky 
boundary conditions are imposed, such as at $\mathscr{I}^+$ in 
asymptotically flat spacetimes.  Since the CLP charges agree with the 
standard expression (\ref{eqn:Hxi}) for charges constructed by
a Wald-Zoupas procedure, when applied in the context of, for example,
$4$D asymptotically flat space, they should reproduce the standard 
expressions for BMS charges and their generalizations 
\cite{Wald:1999wa, Campiglia:2015yka, Campiglia:2014yka, FN, Compere:2018ylh, Flanagan:2019vbl, freidel2021weyl}.
These charges could still fail to satisfy Hamilton's equation
due to the obstruction terms appearing in (\ref{eqn:ham+obs}); however,
this obstruction is fundamentally different from the standard obstruction
appearing in the Wald-Zoupas procedure, which involves the news tensor.
The news tensor should appear in the quantity $\beom_{\ext}$, whereas
the obstruction in (\ref{eqn:ham+obs}) involves the noncovariance 
of the boundary Lagrangian $\ell$ and the corner term $\beta$.  In
particular, $\Delta_\h\xi\ell$ is the quantity that shows up in the 
expression for the extension in the bracket of Wald-Zoupas charges
\cite{Chandrasekaran:2020wwn, Chandrasekaran2021long}, and hence for asymptotic symmetries that produce no such extension,
it is likely that the corresponding CLP charges integrate Hamilton's 
equation for the diffeomorphism.  An interesting future direction would 
be to explore the applications of the CLP extended phase space to 
asymptotic symmetries an to examine the integrability properties 
in more detail.

We found that the charges precisely reproduce the algebra 
satisfied by the vector fields without extension, according to 
(\ref{eqn:HxizetaBT}).  However, in some contexts, the extension term
in the symmetry generators yields important information about the 
theory.  For example, the Brown-Henneaux central extension in the asymptotic
symmetries of $\text{AdS}_3$ 
\cite{Brown-Henneaux} determines the central charge of the dual CFT,
and extensions of symmetries on black hole horizons can in some cases 
provide a derivation of the black hole entropy 
\cite{Strominger1998, Carlip_1999, Hawking_2016,Haco:2018ske,Chen:2020nyh,
Chandrasekaran:2020wwn}.  
The question then arises as to whether any such extension terms 
can appear in the CLP extended phase space.  Perhaps they can arise due to a 
difference between the BT bracket (\ref{eqn:BTbrack}) and the Poisson bracket 
of the charges, due to some noncommutativity of the flux terms in (\ref{eqn:pbflx}).

One interpretation of the results on black hole entropy is that the extra
degrees of freedom associated with the extended phase space yield a 
contribution to the entropy of the region outside the horizon.  
One of the primary motivations for introducing the extended phase space 
is to attempt to give a definition of entanglement entropy in gravity,
which is complicated by the lack of factorization of the classical
phase space due to gauge constraints.  In this picture, one seeks to 
construct the global phase space via a gluing procedure of two extended 
phase spaces associated with a subregion and its complement
\cite{Donnelly2016a}.  This gluing procedure involves a symplectic 
reduction of the product of the two extended phase space in which the 
the charges are matched at the boundary to produce zero total charge, which
ensures the gauge constraints hold on the global phase space.  This same
procedure should continue to hold in the CLP extended phase space.  In fact,
the CLP construction offers several advantages since the charges represent 
the diffeomorphism algebra of the vector fields without extension, and 
hence lead to a more direct application of the symplectic reduction
procedure.  On the other hand, it has been speculated that the existence
of central extensions in the charge algebra could be indicative of a 
reduction procedure that does not fully eliminate all the extended 
degrees of freedom, and the entropy of the subregion could related 
to the leftover degrees of freedom that are not eliminated during the 
reduction \cite{Chandrasekaran2021long}.  It would be interesting how this 
conjecture plays out for the CLP extended phase space, whose charge algebra
never exhibits extension terms.  Instead, the object that would serve as 
an extension appears as an obstruction to integrability of the 
ambiguity-free charges, and it would be interesting to examine the 
effects of this obstruction to the symplectic reduction procedure.  

\section*{Acknowledgments}
I thank Rob Leigh for helpful discussions.  This 
work is supported by the Air Force Office of Scientific Research under award number FA9550-19-1-036.

\bibliographystyle{JHEPthesis}
\bibliography{surfacecharges}

\end{document}